\documentclass[12pt]{iopart}
\usepackage{graphicx}
\pdfoutput=1
\usepackage[normalem]{ulem}
\usepackage[dvipsnames]{xcolor}
\usepackage[utf8]{inputenc}
\usepackage[T1]{fontenc}
\usepackage{mathptmx}
\usepackage{dcolumn}
\usepackage{bm}
\usepackage{textcomp}
\usepackage{caption}
\renewcommand{\vec}[1]{\boldsymbol{#1}}
\begin{document}
\title[Superfluid memory]{Memory device employing hysteretic properties of a tungsten filament in superfluid helium-4}
\author{Che-Chi Shih$^1$, Ming-Huei Huang$^1$,
Pang-Chia Chang$^1$,Po-Wei Yu$^1$,Wen-Bin Jian$^1$,Kimitoshi Kono$^2$$^,$$^3$$^,$$^4$}
\address{$^1$ Department of Electrophysics, National Chiao Tung
  University, Hsinchu 300, Taiwan}
\address{$^2$ International College of Semiconductor Technology, National
  Chiao Tung University, Hsinchu 300, Taiwan}
\address{$^3$ RIKEN CEMS, Hirosawa 2-1, Wako-shi, 351-0198 Japan}
\address{$^4$ Institute of Physics, Kazan Federal University, Kazan,
  420008 Russia}
\ead{a271711@hotmail.com,kkono@nctu.edu.tw.}

\begin{abstract}
  A tungsten filament immersed in superfluid helium has strong
  hysteretic $I$-$V$ characteristics.  By increasing the applied voltage,
  a remarkable current drop occurs at a transition voltage, at which the
  filament enters a non-ohmic hot state and becomes covered with a helium gas
  sheath.  The return to an ohmic state occurs at a lower voltage
  because of the poor heat conduction of the gas sheath. Hence, the
  $I$-$V$ characteristic is strongly hysteretic.  The stable hysteresis
  window was employed to fabricate a novel memory device, which
  demonstrated fast switching and stable reading.
\end{abstract}

\section{Introduction}

Hysteresis is essential to the operation of memory
devices.\cite{waser2010nanoionics,sun2019unified} In recent decades, a
number of novel mechanisms of hysteresis have been explored to realize
new memory devices.\cite{kang2019high,xia2019memristive} An extremely
strong nonlinear electric transport by a tungsten filament immersed in
superfluid helium-4 ($^4$He) was reported by Date \emph{et al.{}} some
time ago.\cite{Date1973} They observed hysteresis in the $I$-$V$
characteristics.  The phenomenon was attributed to the formation of a
He gas sheath surrounding the filament.  The gas sheath is difficult
to produce because of the high thermal conductivity of superfluid
helium,\cite{Pobell1996MatterMethod} but once formed it is difficult
to collapse because of the poor thermal conductivity of the gas
sheath.  Actually, a glowing tungsten filament inside superfluid
helium is one of the most popular demonstrations of superfluid helium.
The filament glows as if it were in a vacuum.  The liquid does not
boil around the filament because the heat produced by the filament is
immediately transported to a free surface and lost via evaporation.
Paradoxically, however, because the thermal conductivity of a
superfluid is infinitely high, the filament is cooled so efficiently
that it is difficult to increase the temperature sufficiently for it
to start glowing.  This is resolved by the formation of a gas sheath
wrapped around the
filament.\cite{Andronikashvili1956,Vinson1968,Okuda1973}

To understand the threshold of gas sheath formation and hysteresis, it
is useful to know some basic properties of superfluid He.  First, it
has a high thermal conductivity because the heat is transported by the
dynamic fluid flow in superfluid $^4$He instead of the usual
diffusion.  A superfluid consists of a macroscopic quantum condensate
and an ensemble of thermally excited elementary excitations, namely
phonons and rotons.  The condensate is in the macroscopic quantum
ground state and hence, carry no entropy, whereas the elementary
excitations carry entropy and heat. The hydrodynamic motion of the
condensate is characterized by the velocity field, $\vec{v}_s$, and
its density $\rho_s$, and referred to as a superfluid component.  The
ensemble of elementary excitations in the condensate acts like a
normal fluid. Its density is defined by $\rho_n = \rho - \rho_s$ and
the flow field is $\vec{v}_n$, where $\rho$ is the total density of
liquid helium.  The total flux of the flow is
$\vec{j} = \rho_s\vec{v}_s + \rho_n\vec{v}_n$.  This formulation is
due to Landau's two-fluid model.\cite{Landau1979Fluid} The heat flux
$\vec{q}$ is then given by
\begin{center}
\begin{equation}
  \label{eq:heat-flux}
 \vec{q} = \rho{s}T\vec{v}_n,
\end{equation}
\end{center}
where $s$ is the entropy per unit mass and $T$ is temperature.  The
thermal conductivity is limited by the effective friction force between
the two fluids.  It is understood that the friction is mediated by
quantized vortex filaments in the superfluid component.\cite{Vinen1957c} In
practice, the thermal conductivity of superfluid helium can be
regarded as infinite.  In this regime, the superfluid $^4$He touches the filament surface and the heat produced is efficiently removed from the filament, so that the filament stays cold and its temperature is uniform along its axis.

\begin{figure}
\begin{center}
\includegraphics{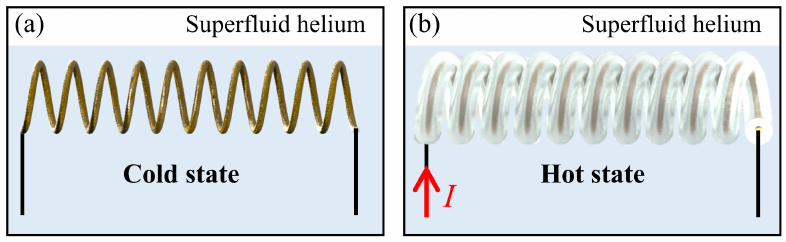}
\caption{\label{fig:schematic} (a) Schematic for the cold state of the filament in superfluid helium. 
(b) Schematic for the hot state of the filament covered with helium gas sheath in superfluid helium.}
\end{center}
\end{figure}

As more heat is produced, however, a gas sheath may appear. This
occurs when the superfluidity is destroyed either by hitting a
liquid--vapor phase boundary in the phase diagram of $^4$He (Refs.~
\cite{Date1973} and \cite{Vinson1968}) or because other critical
conditions are exceeded.  Accordingly, we should expect a transition
from the cold state ($C$-state), where the superfluid is attached to
the filament, to the hot state ($H$-state), in which the filament is
covered by the gas sheath. Schematic drawings of the $C$- and
$H$-states, which are associated with hysteresis, are shown in
Figs.~\ref{fig:schematic}(a) and \ref{fig:schematic}(b). These two
states are distinguished by their different filament resistance
values.  This transition gives rise to the hysteretic transport
properties of the tungsten filament in superfluid $^4$He.  In this
study, we experimentally explored the electric characteristics of a
tungsten filament in superfluid $^4$He and utilized its hysteresis in
the current versus voltage ($I$-$V$) characteristics to realize a
memory device.  This work must open a new research frontier in nano
superfluidics and nano scale gas-liquid phase transitions.

\section{Experimental methods}

\begin{figure}
\begin{center}
\includegraphics{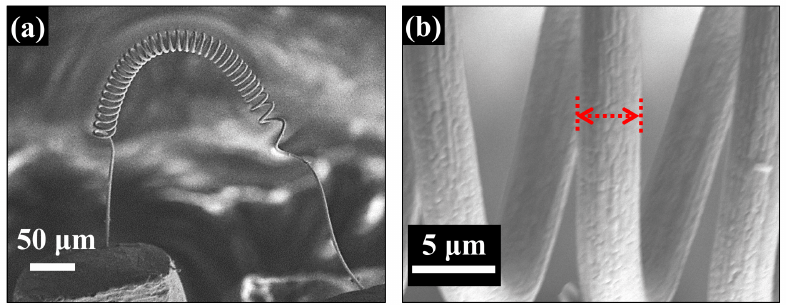}
\caption{\label{fig:filament} (a) SEM image of
  a coiled tungsten filament (sample F-01). (b) Magnified SEM image of
  tungsten wire (sample F-01).}
\end{center}
\end{figure}

The filaments were prepared by removing the glass covers from miniature light bulbs
(Micro-Gl{\"{u}}hlampen-Gesellschaft, Hamburg, Germany). Figures~\ref{fig:filament}(a) and \ref{fig:filament}(b) show scanning
electron microscope (SEM; JEOL IT-300) images of a typical filament
(sample F-01), which was coiled and anchored at two metal
electrodes. To determine the geometrical parameters of the two types
of filament, we obtained SEM images of several filament samples from
the same batch. The average radius of the filament wire and the
diameter of the coil were measured. The total length of the filament
was calculated from the number of coil turns and the length of the
uncoiled part of the filament at both ends. The average total surface
area of the filament was estimated from these size parameters. The surface area of two types of filament were measured and listed in Table~\ref{tab:sample}.

The $I$-$V$ characteristics were studied with four-wire measurements
using a Keithley 2400 source meter. The voltage was swept in a 0.002\,V step with a waiting time of 100\,ms for each step. The experimental setup consisted of liquid helium in a cylindrical glass Dewar surrounded by liquid
nitrogen. The temperature was controlled by regulating the saturated
vapor pressure of the liquid helium. The measurements were performed
in a temperature range below the superfluid transition temperature
$T_\lambda$~(2.1768\,K) of liquid $^4$He.

\begin{table*}
  \caption{\label{tab:sample}Sizes of
    tungsten filament samples.}
\begin{center}
\begin{footnotesize}
\begin{tabular}{cccccc}
\br
 Filament type & Model number & Diameter (\textmu{}m) & Length (mm) & Coil radius (\textmu{}m) & Surface area (mm$^2$)\\ \hline
 F-01 & 2012-00 & $3.97 \pm 0.03$ & $3.06 \pm 0.07$ & $12.26 \pm 0.49$ & $0.0382 \pm 0.0008$ \\
 F-02 & 4032-00 & $9.57 \pm 0.13$ & $4.27 \pm 0.12$ & $24.59 \pm 0.45$ & $0.1284 \pm 0.0053$ \\
\br
\end{tabular}
\end{footnotesize}
\end{center}
\end{table*}

\section{Results and discussion}

\begin{figure}
\begin{center}
\includegraphics{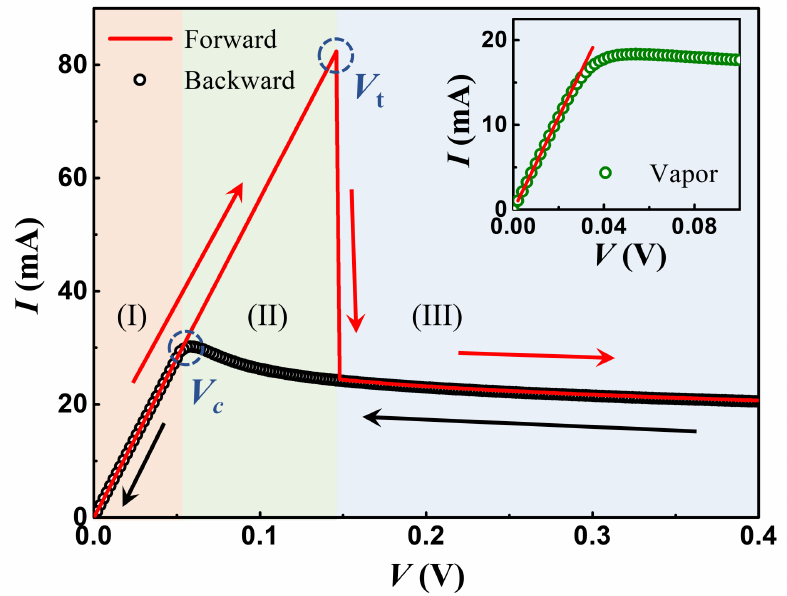}
\caption{\label{fig:iv} Current versus voltage ($I$-$V$)
  characteristics of a tungsten filament (sample F-01) in superfluid
  helium. The red line indicates the forward sweep of the voltage from
  0 to 0.4\,V, and the black open circles indicate the backward sweep
  of the voltage from 0.4 to 0 \,V. $V_t$ is the transition voltage in
  the forward sweep at which the filament forms a sheath and enters
  the $H$-state. $V_c$ is the critical voltage at which the filament
  cools down and enters the ohmic $C$-state. The inset shows the
  $I$-$V$ characteristics of the tungsten filament in the vapor
  phase. A red line is a linear fitting in the low voltage region.}
\end{center}
\end{figure}

Figure~\ref{fig:iv} shows the $I$-$V$ characteristics of the tungsten
filament (sample F-01) with forward and backward sweeps between 0 and
0.4\,V at 1.86\,K. The transition voltage ($V_t$) and critical voltage
($V_c$) divide the behavior into three regions. In the low-power and
intermediate regions I and II of the forward sweep, the filament shows
linear ohmic behavior. The resistance of the filament within these
regions is almost constant at about 1.8\,$\Omega$, corresponding to
the filament temperature.  Due to the high thermal conductivity of a
superfluid, the filament is unable to warm up before reaching
$V_t= 0.146$\,V. In contrast, for a tungsten filament placed over
liquid He in the vapor phase, the $I$-$V$ characteristics are linear
only below 0.02\,V (inset of Fig.~\ref{fig:iv}), which is much lower
than $V_t$. At the transition voltage, a sheath of helium gas
immediately forms along the filament. The sheath is a good thermal
insulation layer, resulting in an abrupt increase of the filament
temperature and resistance.  In the high-power region III, the
temperature and resistance of the filament further increase
efficiently as the voltage increases. Note that the heating process
has a differential negative resistance.\cite{Silvera2008}

In the backward sweep, the device cools as the bias voltage
decreases. The filament is in the $H$-state in regions III and II of
the backward sweep due to the gas sheath. As the voltage decreases,
the sheath shrinks smoothly and eventually disappears at
$V_c = 0.052$\,V. At voltages below $V_c$, the filament returns to an
ohmic state. In regions I and III, the $I$-$V$ characteristics of the
backward sweep completely overlap those of the forward sweep. The
difference between the state transition voltages for the forward and
backward sweeps results in strong hysteresis in region II. The
hysteresis window is defined as the difference between $V_t$ and
$V_c$, which is about 0.094\,V. The hysteresis loop is stable and
reproducible if the environment temperature stays constant. The
hysteresis loop has a promising potential for memory applications.

The relation between the heat conducted to the
liquid and the heat flux produced by the filament is important. As the
voltage increases beyond $V_t$, the superfluid $^4$He near the
filament cannot sustain the direct contact with the filament
surface. A sheath forms due to rapid nucleation. Once a vapor bubble
emerges in some area of the filament surface, the temperature of this
area will immediately increase because of the low thermal conductivity
of helium vapor.  Then, the vapor bubble will rapidly expand to cover
the whole filament.  Unlike a hot filament in a normal liquid, which
boils hard, in superfluid He, the filament is surrounded by a stable
sheath and there is no boiling bubble. Helium gas inside the sheath transports
heat from the surface of the filament to the liquid.

The heat flux density plays an important role in the transition from
the $C$- to the $H$-state and the hysteresis. A superfluid component
carries neither entropy nor viscosity. Therefore, only the normal
component is associated with the heat flux
[Eq.~(\ref{eq:heat-flux})]. We observed that the transition voltage,
$V_t$, is temperature dependent. As the temperature decreases, the
transition voltage appears and increases. It saturates at about
0.14\,V below a temperature of 1.86\,K. Our observations are
qualitatively consistent with previous
reports.\cite{Vinson1968,Date1973,Moss1965} Note, however, that the
critical heat flux densities found are larger than in the previous
work.

\begin{figure}
\begin{center}
\includegraphics{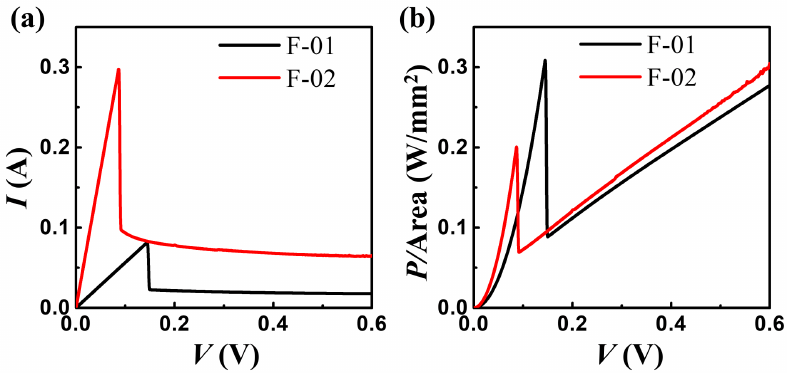}
\caption{\label{fig:power} (a) $I$-$V$ characteristics of tungsten
  filaments, F-01 and F-02 (Table~\ref{tab:sample}), with different
  diameters in superfluid helium.  (b) Heat flux density emitted from
  the filaments as a function of voltage. $T=1.744$\,K.}
\end{center}
\end{figure}

To investigate the effect of heat flux density, two different types of
filament with different diameters were investigated, F-01 and
F-02. Figure~\ref{fig:power}(a) shows the $I$-$V$ characteristics for
the two different samples at 1.744\,K. The thicker filament sample,
F-02, also has a transition. In the ohmic state region, owing to the
ratio of the total length and cross-sectional area, the sample F-02
has a lower resistance of 0.29\,$\Omega$ compared with sample
F-01. The heat flux density as a function of the voltage of the
tungsten filaments is shown in Fig.~\ref{fig:power}(b). Although the
critical currents shown in Fig.~\ref{fig:power}(a) differ, the
critical heat flux densities are close to each other, as shown in
Fig.~\ref{fig:power}(b).  The transition heat flux densities of
samples F-01 and F-02 were 0.316\,W/mm$^2$ (31.6\,W/cm$^2$) and
0.208\,W/mm$^2$ (20.8\,W/cm$^2$), respectively. These values are an
order of magnitude larger than the previously reported values. We
attribute this difference to the preparation of the samples. The
formerly reported values were obtained by employing either Nichrome or
tungsten wire manually fixed to the sample cell. Our tungsten
filaments were from commercial light bulbs. The glass was broken
immediately before the experiment, so that the surfaces should have
been fresh and relatively uncontaminated. Note that the first voltage
scan after a filament was immersed in liquid helium had unreproducible
irregularities, steps, or spikes. The trace became stable and
reproducible only after the first sweep.\cite{Date1973}

From the critical heat flux density of both samples, the velocities of
the normal fluid and the superfluid component can be calculated using
Eq.~(\ref{eq:heat-flux}). $v_n$ and $v_s$ of sample F-01 (F-02) were
2.70~(1.80) and 0.98~(0.65)\,m/s. The critical relative
counterflow velocities, $|\vec{v}_n - \vec{v}_s|$, for F-01 and F-02
were 3.68 and 2.45\,m/s, respectively, at 1.744\,K. Note that
the critical velocity obtained is much larger than the usual critical
velocity for quantum
turbulence.\cite{Vinen1957a,Vinen1957b,Marakov2015,Gao2017,Varga2018}

These large critical values are essential for realizing a stable and wide
hysteresis window.  Moreover, we observed no depth dependence, which
had been reported previously.\cite{Vinson1968,Date1973} The
details of these critical behaviors will be discussed elsewhere.

A memory device is a particularly interesting application of a
filament in superfluid $^4$He. There are two memory states within the
hysteresis window, which can be utilized as program and erase states,
respectively. The hysteresis window is large enough to define the read
voltage for a memory device. Therefore, we demonstrated the memory
functionality with a filament via the same measurement setup.

\begin{figure}
\begin{center}
\includegraphics{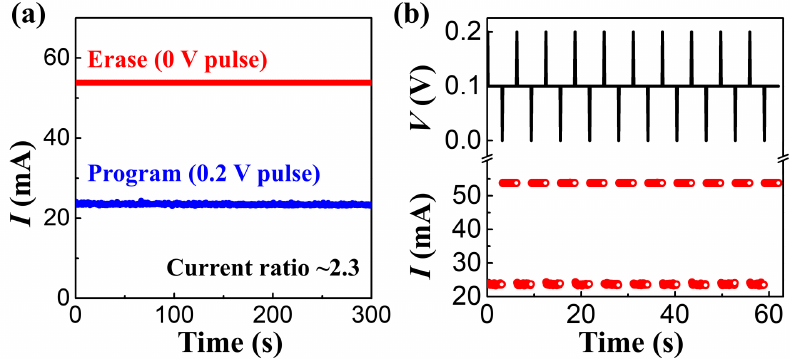}
\caption{\label{fig:memory} (a) Current stability of program and erase
  states of a filament (sample F-01) during 300\,s of hold time. The
  reading voltage was 0.1\,V, and programming and erasing were done at
  0.2\,V for 100\,ms duration and at 0\,V for 100\,ms duration,
  respectively. (b) Reproducible memory operation with 3\,s hold time
  for the same reading and programming conditions as in (a).}
\end{center}
\end{figure}

First, the program and erase voltages were chosen as 0.2\,V and 0\,V,
respectively. The read voltage was defined as 0.1\,V in the middle of
the hysteresis window. In Fig.~\ref{fig:memory}(a), to elucidate the
retention ability of the memory device, program and erase pulses of
duration 100\,ms were applied to filament F-01. Because the erase
state occurs before the formation of the sheath, the mechanism for
reading the current is due to ohmic behavior, which is proportional to
the read voltage.  After applying a pulse of 0.2\,V, the filament
became surrounded by a sheath at the same read voltage, which is
defined as the program state. The filament is hot in the program
state, and the current is remarkably stable. We measured the current
of the two memory states with a read voltage for at least 300\,s. As a
result of the stability of both states, the current ratio was
maintained at about 2.3 for at least 300\,s.  Notice that our device
can switch states without any charge time, which is usually observed
in a charge-trap memory
device.\cite{zhang2015tunable,liu2017eliminating} The memory device
was dynamically switched between the program and erase states by
repeating the erase or program pulse, as shown in
Fig.~\ref{fig:memory}(b). After 10 cycles of programming and erasing,
the two states were still stable, which indicates the reproducibility
and endurance of this memory device.

In addition to the reproducibility and endurance, the writing speed is
another measure of the performance of a memory device.  In a speed
test, however, the time resolution of a four-wire measurement is
limited by the source meter, which is only about 100\,ms. To determine
the shortest pulse width of the program and erase processes, we
employed a function generator (Agilent, 33220a) and an oscilloscope
(Teledyne Lecroy, HDO4034) to measure the dynamic response of the
memory. The experimental setup is shown in Fig.~A1(a) in the Appendix. 
The two-wire hysteresis loop in Fig.~A1(b) in the Appendix is different from that for the four-wire
measurement because of the in-series contact and wire
resistance. Thus, the program, erase, and read voltages were defined
to be 0.5, 0, and 0.25\,V, respectively. The dynamic responses of the
shortest pulse for each process are presented in Figs.~A1(c) and (d)
in the Appendix. The program and erase processes are
attributed to different switching mechanisms, that is, sheath
nucleation and filament cooling. Therefore, the transition time for
programming (300\,\textmu{}s) is shorter than for erasing
(10\,ms). The shortest pulse for the program and erase states are
faster than is usual for a dielectric oxide memory device made from 2D
materials.\cite{zhang2015tunable,liu2017eliminating,lee2015chemically}

\section{Summary and conclusion}

In summary, the results presented in this paper show that a dramatic
current drop is observed during voltage sweeping of a tungsten
filament in superfluid $^4$He. The $C$- and the $H$-states can be
determined from the different voltage bias regions. The state
transitions are due to the formation and annihilation of a helium gas
sheath. The state transitions in the forward and backward sweeps occur
at the transition voltage $V_t$ and critical voltage $V_c$,
respectively. During the voltage sweeping, there is a hysteresis window of about 0.094\,V due to the difference between $V_t$ and
$V_c$. It is suggested that the formation of the sheath depends on a
critical power density. The first-order phase transition occurs at the
critical heat flux from the filament surface.  By defining program,
erase, and read voltages at 0.2, 0, and 0.1\,V, respectively, we
demonstrated the operation of a stable memory device with two
states. The memory device had a strong retention ability, as the
current for both states can remain constant for at least 300\,s. Good
reproducibility and endurance were observed by dynamically switching
between the program and erase states. Our results confirm that the
heating and cooling of a filament produce a stable hysteresis loop in
superfluid $^4$He. A memory device was realized, demonstrating the
promising potential of semiconducting nanowires.

\ack We are indebted to Professor~Ming-Chiang Chung for many
suggestions and comments.  We are grateful to Professor~Ben-Li Young
for helping to set up the cryogenic apparatus, and KK thanks
Professor~Jenh-Yih Juang for his kind hospitality.  This work is
supported by the Ministry of Science and Technology, Taiwan, ROC,
under Grant No. MOST 108-2122-M-009-013.  KK is supported by JSPS
KAKENHI Grant Number JP17H01145 and the Program of Competitive Growth
of Kazan Federal University.

\appendix

\section{Dynamic behavior of the memory effect}

\begin{figure}
\begin{center}
\includegraphics{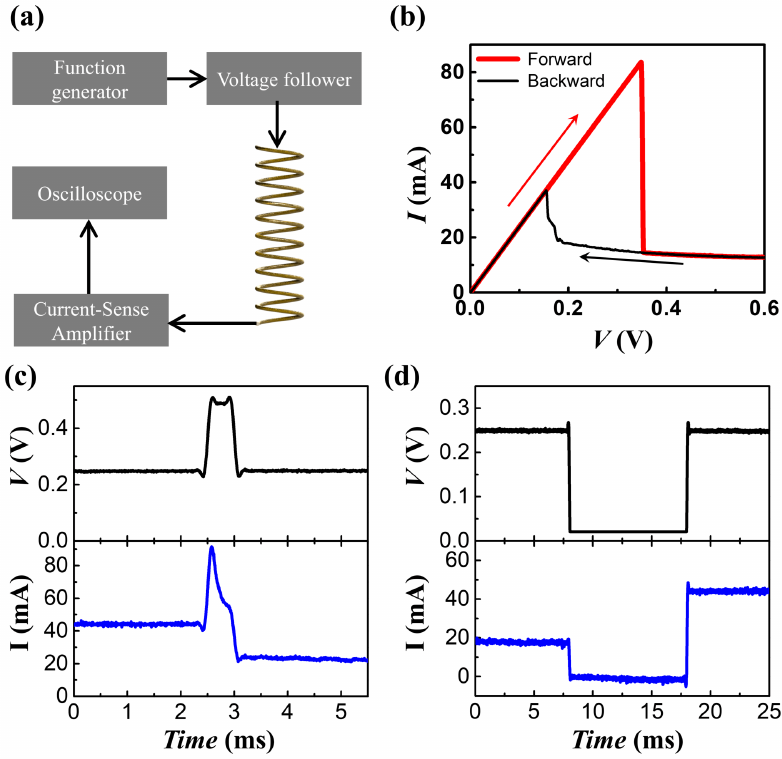}
\caption{\label{fig:dynamic} (a) Diagram of the setup for measuring the
  dynamic behavior. (b) Current versus voltage ($I$-$V$)
  characteristics of a tungsten filament (sample F-01) recorded by
  two-wire measurements. (c) Dynamic response of programming with a
  pulse of 0.5\,V for 300\,\textmu{}s. (d) Dynamic response of the
  erase process with a pulse of 0\,V for 10\,ms.}
\end{center}
\end{figure}

To explore the transition speed of the program and erase states, we
investigated the dynamic behavior of a filament using two-wire current
measurements, as shown in Fig.~\ref{fig:dynamic}(a). A function generator
(Agilent, 33220a) provided signals with different pulse widths. A
voltage follower was connected to buffer the output signal because the
filament impedance is much smaller than 50\,$\Omega$.  The output was
applied to the filament sample, and the current generated was
registered by a current-sense amplifier. An oscilloscope (Teledyne
Lecroy, HDO4034) was used to record the current signal. Due to the
in-series contact and wire resistance, the hysteresis loop in two-wire
measurements is different from that in four-wire measurements, as
described in the main text.

Figure~\ref{fig:dynamic}(b) shows the $I$-$V$ characteristics of the
tungsten filament (sample F-01) recorded by two-wire measurements. The
transition voltage and critical voltage are 0.35\,V and 0.16\,V,
respectively. Hence, to demonstrate the memory behavior in the
two-wire measurements, we defined the program, erase, and read
voltages to be 0.5, 0, and 0.25\,V, respectively.
Figure~\ref{fig:dynamic}(c) shows the dynamic response of the program
process. The reading current was switched from the cold state ($C$-state) to the
hot state ($H$-state) by a program pulse of 0.5\,V in height and 300\,\textmu{}s
in duration.  In our observations, the speed limit of sheath formation
is about 300\,\textmu{}s, which is related to the time constant of
nucleation. To transition the memory from the erase to the program
state, the program pulse of a filament memory device must be equal to
or greater than 300\,\textmu{}s. On the other hand, Fig.~\ref{fig:dynamic}(d)
presents the shortest response time of the dynamic erase process for a pre-programed device. The reading current was switched from the $H$-state to the $C$-state by
an erase pulse with a 0\,V output voltage for 10\,ms. This indicates
that the cooling for the erase state takes about 10\,ms. Hence, the
operating speed of a filament memory device is limited by the cooling
of the erase process.
\section*{Conflicts of interest}
The authors declare no conflict of interest

\section*{References}
\bibliographystyle{iopart-num}
\bibliography{reference}

\end{document}